\documentclass{emulateapj} 
\newcommand{\bl}[1]{\mbox{\boldmath$ #1 $}}

\slugcomment{Accepted for publication in Astrophysical Journal}
\shorttitle{Variable accretion}
\shortauthors{Vorobyov} 

\begin{document}

\title{Variable accretion in the embedded phase of stellar evolution}
\author{E. I. Vorobyov\altaffilmark{1,}\altaffilmark{2}}
\altaffiltext{1}{Institute for Computational Astrophysics, Saint Mary's University,
Halifax, B3H 3C3, Canada; vorobyov@ap.smu.ca.} 
\altaffiltext{2}{Institute of Physics, South Federal University, Stachki 194, Rostov-on-Don, 
344090, Russia.} 

%\maketitle

\begin{abstract}
Motivated by the recent detection of a large number of embedded young stellar objects (YSOs)
with mass accretion rates that are inconsistent with the predictions of the standard model
of inside-out collapse \citep{Shu77},
we perform a series on numerical hydrodynamic simulations of the gravitational collapse 
of molecular cloud cores with various initial masses, rotation rates, and sizes.
We focus on the early Class I stage of stellar evolution when circumstellar disks
are exposed to high rates of mass deposition from infalling envelopes.
Our numerical modeling reproduces the large observed spread in accretion rates 
inferred for embedded YSOs in Perseus, Serpens, and Ophiuchus star forming regions by \citet{Enoch09},
yielding $37\%$--$75\%$ of objects with ``sub-Shu'' accretion rates 
$\dot{M}\la 10^{-6}~M_\odot$~yr$^{-1}$ and $1\%$--$2\%$ of objects with 
``super-Shu'' accretion rates $\dot{M}>10^{-5}~M_\odot$~yr$^{-1}$. Mass accretion
rates in the Class I stage have a log-normal distribution, with its shape controlled
by disk viscosity and disk temperature. The spread in $\dot{M}$ is greater in models with 
lower viscosity and smaller in models with higher viscosity and higher disk temperature,
suggesting that gravitational instability may be a dominant cause of the
observed diversity in $\dot{M}$ in embedded YSOs. Our modeling predicts a weak dependence
between the mean mass accretion rates and stellar masses in the Class I stage, 
in sharp contrast to the corresponding steep dependence for evolved T Tauri stars and
brown dwarfs.

\end{abstract}


\section{Introduction}
It has recently become evident that young stellar objects (YSOs) in the embedded
phase of stellar evolution exhibit a variety of mass accretion rates, which range
between $\dot{M} \sim 10^{-8}~M_\odot~\mathrm{yr}^{-1}$ and  $\dot{M} \sim 
10^{-5}~M_\odot~\mathrm{yr}^{-1}$ \citep[e.g.][]{Dunham06,Enoch09} and 
often cannot be accounted for in the standard model of inside-out collapse \citep{Shu77}.
According to this model, the infall rate of gas from an envelope onto a star/disk system is 
$\dot{M}_{\rm infall}\approx c_{s}^3/G$, where $c_{s}$ is the sound speed.
For the typical gas temperatures in star forming regions $T_{\rm g}$=10--20~K, 
the standard model predicts infall rates of order 1.2--4.2$\times 10^{-6}~M_\odot~\mathrm{yr}^{-1}$.
According to \citet{Enoch09}, however, more than $50\%$ of YSOs in Perseus,
Serpens, and Ophiuchus star-forming regions have inferred  mass accretion rates
$ 10^{-8}~M_\odot~\mathrm{yr}^{-1} \la \dot{M} \la 10^{-6}~M_\odot$~yr$^{-1}$, which are
considerably lower than the infall rates.  At the same time, the envelope mass decreases 
about as predicted by the standard model \citep{Enoch09}, implying that
the observationally inferred and theoretically derived infall rates $\dot{M}_{\rm infall}$ 
agree with each other.

The obvious mismatch between $\dot{M}$ and $\dot{M}_{\rm infall}$ was first noticed by 
\citet{Kenyon90} based on the comparison between typical lifetimes of embedded YSOs
and inferred bolometric luminosities---the accretion luminosity 
$\sim 10^{-6}-10^{-5}~M_\odot$~yr$^{-1}$ inferred from the duration of the embedded phase 
$\mathrm{a~few}\times 10^5$~yr turned out to be much larger than the typical bolometric 
luminosity $\sim 10^{-7}~M_\odot$~yr$^{-1}$. This ``luminosity problem'' can be solved by
the infalling material first piling up in a circumstellar disk and then 
being accreted onto a protostar mostly in short-lived FU-Ori-like episodes with accretion rates 
$\dot{M} > 10^{-5}~M_\odot~\mathrm{yr}^{-1}$. Numerous mechanisms for this episodic 
accretion have 
been proposed in the past. These include the thermal ionization instability 
\citep{Lin85,Hartmann85,Bell94}, close encounters in binary systems \citep{Bonnell92},
gravitational instability and fragmentation in embedded circumstellar disks \citep{VB05,VB06},
a combination the magneto-rotational instability in the inner disk regions 
and gravitational instability in the outer disk regions \citep{Zhu09}, and 
close encounters in young stellar clusters \citep{Pfalzner08}.

Most theoretical and numerical studies of episodic accretion have been focused 
so far on explaining 
observable characteristics of the FU-Ori-like accretion bursts. In this paper,
however, we mostly focus on the quiescent phase of accretion between the bursts.
This phase has recently gained much interest due to a discovery of very 
low luminosity objects or VELLOs in dense molecular cores that have previously been thought of 
as ``starless'' \citep[e.g.][]{Young04,Kauffmann05,Bourke06, Stecklum07}.
One possible explanation for their low luminosity ($L<0.1~L_\odot$) is a quiescent accretion, 
in which $\dot{M}$ has to be more than an order of magnitude lower than the Shu 
infall rates 
1.2--4.2$\times 10^{-6}~M_\odot~\mathrm{yr}^{-1}$, suggesting again that the protostellar 
accretion history may not be as simple as predicted by the standard model.

In this paper, we provide a comprehensive study of the mass accretion 
rates $\dot{M}$ and disk infall rates $\dot{M}_{\rm infall}$ in the embedded phase 
of stellar evolution (focusing mostly on the Class I stage). 
We use numerical hydrodynamic simulations of the cloud core collapse 
with a self-consistent disk formation and long-term evolution. 
We run a large set of model covering a wide range of cloud core masses and angular velocities. 
The general problem of explaining the large spread of observed protostellar 
accretion rates in the Class I stage is addressed using ideas from
the burst mode of accretion discovered recently by \citet{VB05,VB06}.

%The existence of a substantial number of YSOs in the sub-Shu accretion state 
%is a manifestation of the long-known luminosity problem---observations 
%indicate that mass accretion rates onto the star are generally lower 
%than the rate of disk loading from infalling envelopes \citep{Kenyon90,Kenyon94}.

\section{Description of numerical model}
\label{nummodel}

Our numerical model is similar to that used recently by \citet{VB09} to simulate the secular 
evolution of viscous and self-gravitating circumstellar disks 
and the mass accretion rates in the late evolution phase of T Tauri stars (TTSs)  and 
brown dwarfs (BDs) \citep{VB10}. For the reader's convenience, we 
briefly review the basic concept and equations.

We employ numerical hydrodynamic simulations in the thin-disk approximation 
to compute the evolution of rotating, gravitationally unstable cloud cores with 
various initial masses $M_{\rm cl}$ and ratios of rotational to gravitational energy $\beta$. 
We start our numerical integration in the pre-stellar phase, which is 
characterized by a collapsing {\it starless} cloud core, continue into the 
embedded phase of stellar evolution, which sees the formation of a star/disk/envelope
system, and terminate our simulations in the late accretion phase,
when most of the cloud core has accreted onto the star/disk system. 
Once the disk has self-consistently formed, it occupies the innermost 
regions of our numerical grid, while the infalling cloud core (the envelope) 
occupies the rest of the grid. This ensures that the mass infall rate onto 
the disk $\dot{M}_{\rm infall}$ is accurately determined by the dynamics of 
the gas in the envelope. 
%As a result, there is no source term in the numerical 
%grid allowing for some sort of mass deposition from the envelope, but 

%The thin-disk approximation is an excellent means to calculate the evolution
%for many orbital periods and many model parameters. It is well justified as long as
%the aspect ratio of the disk scale height $Z$ to radius $r$ does not considerably 
%exceed 0.1. As one of us has recently shown \citep[see figure 7 in][]{Vor09}, this condition
%is fulfilled for solar mass stars having disks of several hundred AU in radius.
%Our model disks rarely exceed this size, and hence we believe that the thin-disk approximation
%is justified in our modeling.

The basic equations of mass and momentum transport in the thin-disk approximation are
\begin{eqnarray}
\label{cont}
 \frac{{\partial \Sigma }}{{\partial t}} & = & - \nabla _p  \cdot \left( \Sigma \bl{v}_p 
\right), \\ 
\label{mom}
 \Sigma \frac{d \bl{v}_p }{d t}  & = &  - \nabla _p {\cal P}  + \Sigma \, \bl{g}_p + 
 (\nabla \cdot \mathbf{\Pi})_p \, ,
\end{eqnarray}
where $\Sigma$ is the mass surface density, ${\cal P}=\int^{Z}_{-Z} P dz$ is the vertically integrated
form of the gas pressure $P$, $Z$ is the radially and azimuthally varying vertical scale height,
$\bl{v}_p=v_r \hat{\bl r}+ v_\phi \hat{\bl \phi}$ is the velocity in the
disk plane, $\bl{g}_p=g_r \hat{\bl r} +g_\phi \hat{\bl \phi}$ is the gravitational acceleration 
in the disk plane, and $\nabla_p=\hat{\bl r} \partial / \partial r + \hat{\bl \phi} r^{-1} 
\partial / \partial \phi $ is the gradient along the planar coordinates of the disk. 
The gravitational acceleration $\bl{g}_p$ includes both the gravity of a central point object 
(when formed)
and the self-gravity of a circumstellar disk and envelope. The latter component is found 
by solving the Poisson integral using the convolution theorem.
The viscous stress tensor $\mathbf{\Pi}$ is expressed as
\begin{equation}
\mathbf{\Pi}=2 \Sigma\, \nu \left( \nabla v - {1 \over 3} (\nabla \cdot v) \mathbf{e} \right),
\end{equation}
where $\nabla v$ is a symmetrized velocity gradient tensor, $\mathbf{e}$ is the unit tensor, and
$\nu$ is the kinematic viscosity. 
%Equation~(\ref{mom}) describes the motion of a viscous fluid in
%the most general form. This equation can be reduced to the usual equation for the conservation 
%of angular momentum of a radial annulus in the axisymmetric accretion disk.
The components of $(\nabla \cdot \mathbf{\Pi})_p$ in polar coordinates ($r,\phi$) 
can be found in \citet{VB09}. 

%It is well known that standard collisional viscosity (molecular viscosity) 
%is negligible in application to circumstellar disks.
The best candidate for viscosity in circumstellar disks 
is turbulence induced by the magneto-rotational 
instability (MRI), though other mechanisms such as nonlinear hydrodynamic turbulence cannot 
be completely eliminated due to the large Reynolds numbers involved. 
We make no assumptions about the source of viscosity and parameterize its magnitude 
using the usual $\alpha$-prescription \citep{SS}
\begin{equation}
\label{viscosity}
\nu = \alpha \, \tilde{c}_{s} \, Z,
\end{equation}
where $\tilde{c}^2_{s}=\partial {\cal P} /\partial \Sigma$ is the effective sound speed
of (generally) non-isothermal gas. The vertical scale height $Z$ is determined in
every computational cell and at every time step of integration using
an assumption of local hydrostatic equilibrium in the gravitational field of
the central star and the disk \citep[see][]{VB09}. 

For most of the numerical simulations in this paper, we use $\alpha=0.01$. This choice 
is motivated by the recent work
of \citet{VB09}, who studied numerically the secular evolution
of viscous and self-gravitating disks. They found that {\it if} circumstellar disks
around solar-mass protostars could generate and sustain turbulence than the temporally and 
spatially averaged $\alpha$ should lie in the $10^{-3}-10^{-2}$ limits. 
Smaller values of $\alpha$ ($\la 10^{-4}$) have little effect on the resultant mass accretion history,
while larger values ($\alpha \ga 10^{-1}$) destroy circumstellar disks during less than 1.0~Myr of 
evolution and are thus unlikely from the point of view of disk longevity. 
We note that we have intentionally taken the largest possible value for $\alpha$,
since we want to assess the maximum effect that viscosity may have on the accretion 
history in the embedded phase. The case of a smaller $\alpha=10^{-3}$ will 
be considered in brief in Section~\ref{lowervisc}. We also note that viscosity 
is introduced in numerical simulations only after disk formation. In the pre-disk 
phase, the $\alpha$-parameter is set to zero.

%The motivation of the present paper is to explore the effect that viscosity
%may have on the mass accretion rates in the embedded phase of stellar evolution.
%As was recently shown by \citet{VB09}, viscosity moderates the burst phenomenon
%and accretion variability ... By using a wide range of models we want to accurately
%assess the magnitude of this effect...

Equations~(\ref{cont}) and (\ref{mom}) are closed with a barotropic equation
that makes a smooth transition from isothermal to adiabatic evolution at 
$\Sigma = \Sigma_{\rm cr} = 36.2$~g~cm$^{-2}$:
\begin{equation}
{\cal P}=c_s^2 \Sigma +c_s^2 \Sigma_{\rm cr} \left( \Sigma \over \Sigma_{\rm cr} \right)^{\gamma},
\label{barotropic}
\end{equation}
where $c_s=0.188$~km~s$^{-1}$ is the sound speed in the beginning of numerical simulations 
(corresponding to the initial gas temperature of 10~K) and $\gamma=1.4$. 
The adopted value for $\Sigma_{\rm cr}$ corresponds to the gas
volume number density of $10^{-11}$~cm$^{-3}$ for a disk in the vertical hydrostatic equilibrium
at temperature 10~K. The effect of larger $\gamma$ and, as a consequence, hotter circumstellar
disks is explored in Section~\ref{hotdisk}.

Equations~(\ref{cont}) and (\ref{mom}) are solved in polar 
coordinates $(r, \phi)$ on a numerical grid with
$128 \times 128$ points. We have found that an increase in the resolution to 
$256 \times 256$ grid zones makes little influence on the accretion history 
but helps to save a considerable amount of CPU time and consider many more model cloud cores. 
%Each model takes about 300--400 CPU hours on the Opteron 2.5 GHz processor.
We use the method of finite differences with a time-explicit,
operator-split solution procedure similar in methodology to the ZEUS code. Advection is
performed using the second-order van Leer scheme.  The radial points are logarithmically spaced.
The innermost grid point is located at $r_{\rm sc}=5$~AU, and the size of the 
first adjacent cell varies in the 0.17--0.36~AU range depending on the cloud core size.  
We introduce a ``sink cell'' at $r<5$~AU, 
which represents the central star plus some circumstellar disk material, 
and impose a free inflow inner boundary condition. The outer boundary is reflecting.
%The gravity of a thin disk is computed by directly summing the input from each computational cell
%to the total gravitational potential. The convolution theorem is used to speed up 
%the summation. 
A small amount of artificial viscosity is added to the code, 
though the associated artificial viscosity torques are always negligible 
in comparison with the gravitational torques. 

\begin{table*}
\center
\caption{Model parameters}
\label{table1}
%\vspace{3 pt}
\begin{tabular}{ccccccc}
\hline\hline
Set & $\beta$ & $\Omega_0$ & $r_0$ & $r_{\rm out}$ & $M_{\rm cl}$ & N  \\
\hline
 1 & $1.18\times 10^{-3}$ & $0.27-1.00$ & $1380-5180$ & $(0.8-3.0)\times 10^4$  & $0.8-3.0$  & 8   \\
 2 & $2.29\times 10^{-3}$ & $0.37-2.24$ & $860-5180$  & $(0.5-3.0)\times 10^4$  & $0.5-3.0$  & 11   \\
 3 & $3.37\times 10^{-3}$ & $0.45-3.40$ & $690-5180$  & $(0.4-3.0)\times 10^4$  & $0.4-3.0$  & 14   \\
 4 & $5.10\times 10^{-3}$ & $0.53-6.70$ & $410-5180$  & $(0.2-3.0)\times 10^4$  & $0.24-3.0$ & 12  \\
 5 & $7.25\times 10^{-3}$ & $0.67-8.30$ & $414-5180$  & $(0.2-3.0)\times 10^4$  & $0.24-3.0$ & 9 \\
 6 & $1.20\times 10^{-2}$ & $1.20-17.0$ & $240-3450$  & $(0.14-2.0)\times 10^4$ & $0.14-2.0$ & 11 \\
 7 & $1.42\times 10^{-2}$ & $2.30-14.0$ & $340-2070$  & $(0.2-1.2)\times 10^4$ &  $0.2-1.2$  & 4 \\
 8 & $2.00\times 10^{-2}$ & $1.60-22.9$ & $240-3450$  & $(0.14-2.0)\times 10^4$ & $0.14-2.0$ & 5 \\
 9 & $2.90\times 10^{-2}$ & $2.00-20.0$ & $340-3450$   & $(0.2-2.0)\times 10^4$ & $0.2-2.0$  & 4 \\
 \hline
\end{tabular} 
\tablecomments{All distances are in AU, angular
velocities in km~s$^{-1}$~pc$^{-1}$, masses in $M_\odot$, and $N$ is the number of models in each set.}
\end{table*} 

\section{Initial conditions}
We start our numerical simulations from starless cloud cores, which have surface densities 
$\Sigma$ and angular velocities $\Omega$ typical for a collapsing axisymmetric magnetically
supercritical core \citep{Basu}
\begin{equation}
\Sigma={r_0 \Sigma_0 \over \sqrt{r^2+r_0^2}}\:,
\label{dens}
\end{equation}
\begin{equation}
\Omega=2\Omega_0 \left( {r_0\over r}\right)^2 \left[\sqrt{1+\left({r\over r_0}\right)^2
} -1\right],
\end{equation}
where $\Omega_0$ is the central angular velocity, 
$r_0$ is the radius of central near-constant-density plateau defined 
as $r_0 = k c_s^2 /(G\Sigma_0)$ and  $k= \sqrt{2}/\pi$.
These initial profiles are characterized by the important
dimensionless free parameter $\eta \equiv  \Omega_0^2r_0^2/c_s^2$
and have the property 
that the asymptotic ($r \gg r_0$) ratio of centrifugal to gravitational
acceleration has magnitude $\sqrt{2}\,\eta$.
The centrifugal radius of a mass shell initially located at radius $r$ is estimated to be
$r_{\rm cf} = j^2/(Gm) = \sqrt{2}\, \eta r$, where $j=\Omega r^2$ is the specific angular
momentum. We note that $\eta$ is similar in magnitude to the ratio of
rotational to gravitational energy $\beta=E_{\rm rot}/E_{\rm grav}$, where the
rotational and gravitational energies are calculated as
\begin{equation}
E_{\rm rot}= 2 \pi \int \limits_{r_{\rm sc}}^{r_{\rm
out}} r a_{\rm c} \Sigma \, r \, dr, \,\,\,\,\,\
E_{\rm grav}= - 2\pi \int \limits_{r_{\rm sc}}^{\rm r_{\rm out}} r
g_r \Sigma \, r \, dr.
\label{rotgraven}
\end{equation}
Here, $a_{\rm c} = \Omega^2 r$ is the centrifugal acceleration, and $r_{\rm out}$
is the outer cloud core radius. The numerical relationship between the two parameters
is $\beta=0.91 \eta$.
The gas has a mean molecular 
mass $2.33 \, m_{\rm H}$ and cloud cores are initially isothermal with temperature $T=10$~K.

We present results from nine sets of models, the parameters of which are detailed
in Table~\ref{table1}. Each set is characterized by a distinct ratio $\beta$ 
of rotational to gravitational energy. The adopted values of $\beta$ lie within
the limits inferred by \citet{Caselli} for dense molecular cloud cores, $\beta=(10^{-4} - 0.07)$.
%Individual models within every set are generated by varying $r_{\rm out}$, $r_0$, and $\Omega_0$. 
The product $r_0 \Omega_0$ for every model in a given set is kept constant to 
enforce the $\beta=\mathrm{const}$ condition (the initial sound speed 
$c_s = 0.188$ km s$^{-1}$ is equal for all models). 
In addition, the ratio $r_{\rm out}/r_0$ is set to 6.0
to generate gravitationally unstable truncated cores of similar form.
 
The actual procedure for generating a specific cloud core with a given value of $\beta$ 
is as follows. First, we choose
the outer cloud core radius $r_{\rm out}$ and find $r_0$ from the condition $r_{\rm out}/r_0=6.0$.
Then, we find the central surface density $\Sigma_0$ from the relation 
$r_0=\sqrt{2}c_{\rm s}^2/(\pi G \Sigma_0)$ and determine the resulting cloud core mass
$M_{\rm cl}$ from Equation~(\ref{dens}). Finally, the central angular velocity $\Omega_0$
is found from the condition $\beta=0.91\eta=0.91\Omega_0^2 r_0^2/c_{\rm s}^2$. 

In total, we have simulated numerically the time evolution of 78 cloud cores. 
We note that model sets with greater $\beta$ have on general a smaller number
of models due to numerical difficulties associated with modeling of massive cloud cores with
high angular velocities.
The resulting initial cloud core mass function $dN/dM_{\rm cl}=M_{\rm cl}^{m}$ has
exponents $m_1=-0.4\pm0.1$ in the $0.1~M_\odot < M_{\rm cl} < 1.0~M_\odot$ mass range 
and $m_2=-1.3\pm 0.2$ in the $1.0~M_\odot < M_{\rm cl} < 3.0~M_\odot$ range. 
These values are somewhat shallower than usually inferred from nearby star-forming regions
\citep[e.g.][]{Enoch06}. The slope of the cloud core mass function varies 
in different star forming regions and the effect of this variation 
will be considered in a follow-up study. 
% e made no specific attempt to fit our core mass function into 
%any of the observed

%to make our truncated cloud cores resemble Bonnor-Ebert spheres and guarantee that
%they are gravitationally unstable.

%The values of $\beta$, typical intervals for $r_{\rm out}$, $r_0$, $\Omega_0$ 
%and cloud core masses $M_{\rm cl}$, and number of individual models within each set are listed 
%in Table~\ref{table1}. 

\section{Time evolution of mass accretion rates}
\label{timeevol}
In this section, we review the accretion history in the early phase of stellar evolution 
as derived using our numerical hydrodynamic simulations.
We calculate the instantaneous mass accretion rate $\dot{M}=-2 \pi r_{\rm sc} v_{r_{\rm sc}} \Sigma(r_{\rm
sc})$
as the mass passing through the sink cell per one time step of numerical 
integration (which in physical units is usually equal to 10--20 days). 
We note that the size of the sink cell $r_{\rm sc}=5$~AU
is larger than the stellar radius. The inner disk at $r<5$~AU may add 
additional variability to the accretion rates, in particular due to the
thermal ionization instability \citep{Bell94} or MRI \citep{Zhu09}.
These effects may somewhat alter the temporal behaviour of the actual accretion
rates onto the stellar surface. Protostellar jets may also decrease the rate of
mass deposition onto the star by about 10\%.
% and our model accretion rates 
%may not be directly related to the observationally inferred accretion rates. 
We leave these complicated phenomena for a future study.

\begin{figure*}
 \centering
  \includegraphics[width=18cm]{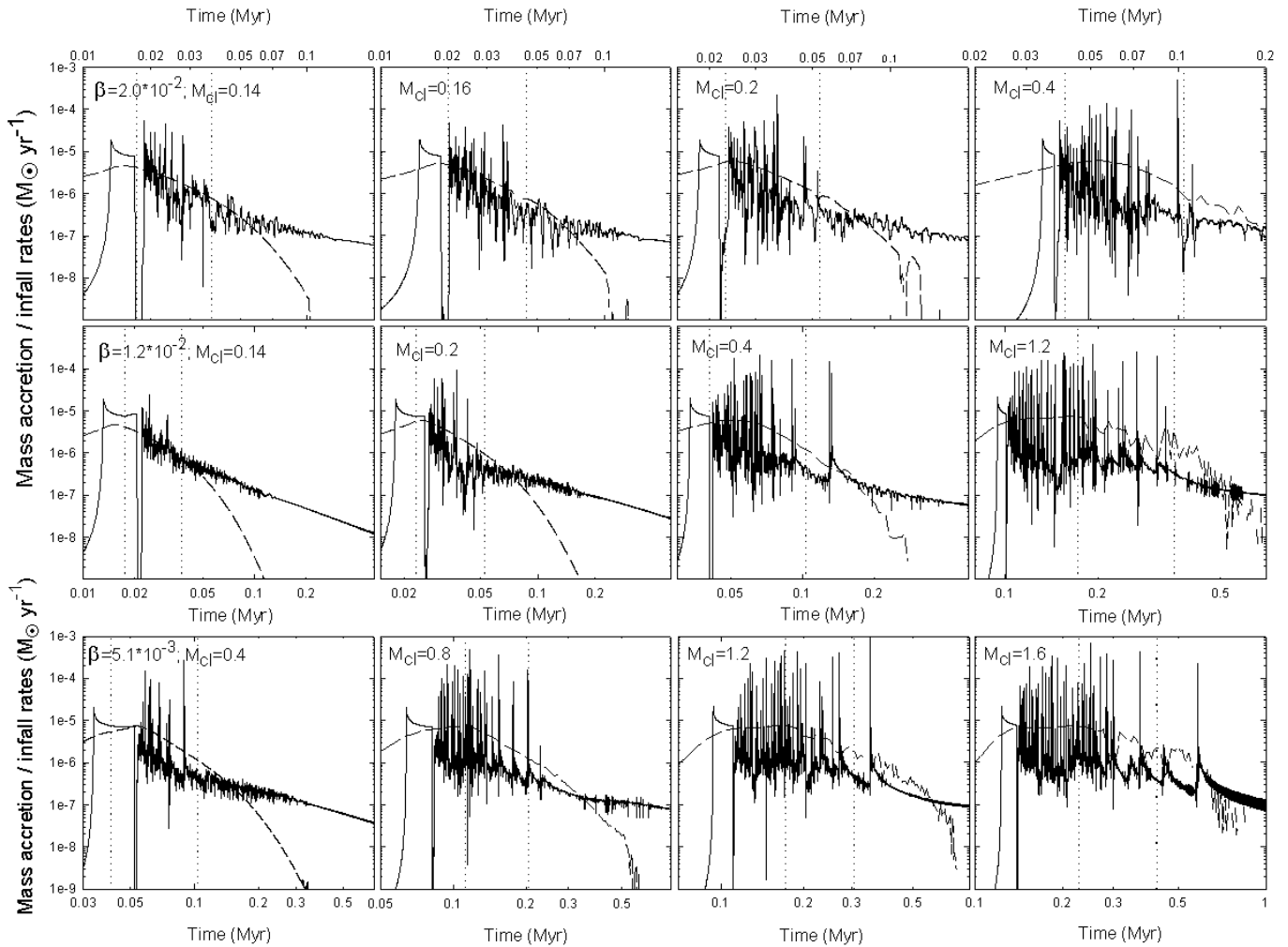}
      \caption{Mass accretion rates $\dot{M}$ (solid lines) and mass infall rates $\dot{M_{\rm infall}}$
       (dashed lines) versus time elapsed since the beginning of numerical
      simulations. Twelve models with distinct initial cloud core masses 
      $M_{\rm cl}$ (in $M_\odot$) and ratios of rotational to gravitational energy $\beta$ 
      are presented. In particular, the top/middle/bottom rows have $\beta=2.0\times 10^{-2}/1.2\times
      10^{-2}/5.1\times 10^{-3}$, respectively. The vertical dotted lines mark 
      the onset of the Class I stage (left) and Class II stage (right) of stellar evolution. }
         \label{fig1}
\end{figure*}

Figure~\ref{fig1} shows the time behaviour of $\dot{M}$ (solid lines)  for 12 models 
characterized by distinct initial cloud core masses $M_{\rm cl}$ (in $M_\odot$) and ratios of the 
rotational to gravitational energy $\beta$. In addition, the dashed lines present
mass infall rates  calculated as $\dot{M}_{\rm infall}= - 2 \pi r v_r \Sigma(r)$, where $r=600$~AU.
The disk radius in our models rarely exceeds 600~AU and $\dot{M}_{\rm infall}$
represents the rate of gas deposition from an infalling envelope onto a disk.
The vertical dotted lines mark the onset of the Class I (left) and Class II (right)
stages of stellar evolution, as inferred from the mass remaining in the envelope. 
Following a prescription of \citet{Andre93} (which is most 
useful for numerical simulations), we assume that the Class I/Class II stages
ensue when 50\%/90\% of the initial cloud core mass, respectively, has been accreted 
onto the star plus disk system. Other classification schemes may introduce a 
systematic shift of order unity.

The early evolution of the mass accretion rate is similar in all models---$\dot{M}$ reaches a maximum
value of $\sim 2\times 10^{-5}~M_\odot$~yr$^{-1}$ and then settles at a near-constant 
value of order of $10^{-5}~M_\odot$~yr$^{-1}$. This early phase of near-constant accretion
ends when the first layer of infalling material hits the centrifugal radius at 5.0~AU and a disk
starts to form in our numerical simulations. The subsequent evolution of mass accretion is controlled
by the physical processes in the disk and the rate of mass loading $\dot{M}_{\rm infall}$ onto 
the disk.
A continuing deposition of matter from the infalling envelope, as indicated by the dashed lines
in Figure~\ref{fig1}, leads to the development of gravitational
instability and fragmentation in the disk. Fragments typically form at $r\ga 40$~AU, 
have typical masses of up to 10--20 Jupiters, sizes of several AU, number densities of up to
$10^{13}-10^{14}$~cm$^{-3}$, and are pressure supported against their own gravity. 
These fragments are quickly driven into the inner disk regions via the gravitational exchange
of angular momentum with the spiral arms and trigger a short-lived burst of mass accretion
when passing through the sink cell. This phenomenon is called the burst mode
of accretion and is investigated in detail by \citet{VB05,VB06}. The ultimate fate of the fragments
is uncertain and is largely dependent on how quickly they can contract from their initial size
of several AU to the planetary size to avoid tidal destruction.
%Provided that they have enough time to dissociate molecular hydrogen, 
%they may form massive hot jupiters at close orbits around the central star.
According to \citet{Helled06}, the contraction time for a Jupiter-mass clump to reach
a central temperature of 2000~K, i.e., the temperature required to dissociate H$_2$ to trigger
rapid collapse, is $3\times10^5$~yr.
However, considering a fast time scale of inward radial migration to the inner 5~AU---a few 
tens of orbital periods at most\footnote{The animation of this process can be 
viewed at www.astro.uwo.ca/$\sim$vorobyov}---we believe that most of these fragments will be tidally destroyed when 
approaching the central star, thus converting its gravitational energy to the accretion luminosity 
and producing an FU-Ori-like luminosity burst. 

Figure~\ref{fig1} demonstrates that the burst phenomenon is more pronounced in models with
greater initial cloud core masses $M_{\rm cl}$.  
Indeed, more massive cloud cores have larger outer radii $r_{\rm out}$ and, as a consequence, 
larger centrifugal radii $r_{\rm cf}$.
%$r_{\rm out}$, which is greater for a cloud core of greater mass. 
%%is constructed by increasing its outer radius $r_{\rm out}$ (but 
%%keeping the ratio $r_{\rm out}/r_0$ constant).
It then follows that models with greater $M_{\rm cl}$ are expected to form more massive disks, 
which are easier susceptible to gravitational instability and fragmentation. Another important 
feature that can be seen in Figure~\ref{fig1}
is that the bursts cease when the rate of mass deposition onto the disk $\dot{M}_{\rm infall}$
(dashed lines) falls below that of the mass accretion $\dot{M}$ (solid lines). This clearly demonstrates
the destabilizing influence that the infalling envelope has on the disk. The burst phenomenon is 
mostly localised to the Class 0 and Class I stages of stellar evolution,
with only a few bursts taking place in the early Class II stage.

%We however believe that if
%there had been a significant mismatch between these values, then there would 
%have been observational signatures such as material piling up in the inner 
%few AU. 

%A particular emphasis is payed to the mass accretion burst phenomenon, 
%originally discovered by \citet{VB05,VB06}.

%{\it Here, a figure showing the instantaneous mass accretion rates should go,
%emphasizing the burst phenomenon and accretion variability. Disk infall rates 
%should also be shown.} 

\section{Accretion variations in the embedded phase of stellar evolution}
%Our numerical modeling yields a highly variable accretion history in
%the phase that immediately follows disk formation. 
%Prolonged periods of rather low accretion $ \dot{M} \la 10^{-6}~M_\odot$~yr$^{-1}$,
%when the disk may be weakly gravitationally unstable but is yet stable to fragmentation,
%are interspersed with short-lived bursts with $\dot{M}\ga 10^{-5}~M_\odot$~yr$^{-1}$, when
%the disk forms fragments and drives them onto the central star.
Accretion variations appear to be an intrinsic property of self-gravitating disks in
the early embedded phase of stellar evolution. Young circumstellar disks, when exposed to mass
deposition from infalling envelopes, undergo repeating cycles of gravitational destabilization
and fragmentation.
Prolonged periods of rather low accretion $ \dot{M} \la 10^{-6}~M_\odot$~yr$^{-1}$,
when the disk may be gravitationally unstable but is yet stable to fragmentation,
are interspersed with short-lived bursts with $\dot{M}\ga 10^{-5}~M_\odot$~yr$^{-1}$, when
the disk forms fragments and drives them onto the central star.
It is therefore interesting to directly compare our predicted spread in $\dot{M}$ 
with that inferred 
in YSOs in Perseus, Serpens, and Ophiuchus star-forming regions  by \citet{Enoch09}.
In the remaining text, we focus on the Class I stage of stellar evolution,
leaving the Class 0 stage for a follow-up study.

\subsection{Distribution Functions of Accretion Rates}

\begin{figure}
  \resizebox{\hsize}{!}{\includegraphics{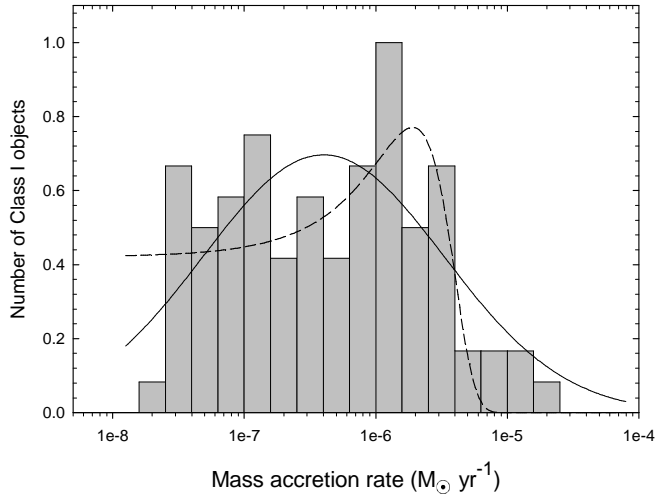}}
      \caption{The normalized number of Class I YSOs versus mass accretion rate as
      inferred by \citet{Enoch09} in Perseus, Serpens and Ophiuchus star forming regions. 
      The solid and dashed lines are the log-normal
      and gaussian fits to the observed data. }
         \label{fig2}
\end{figure}

\begin{figure*}
 \centering
  \includegraphics[width=18cm]{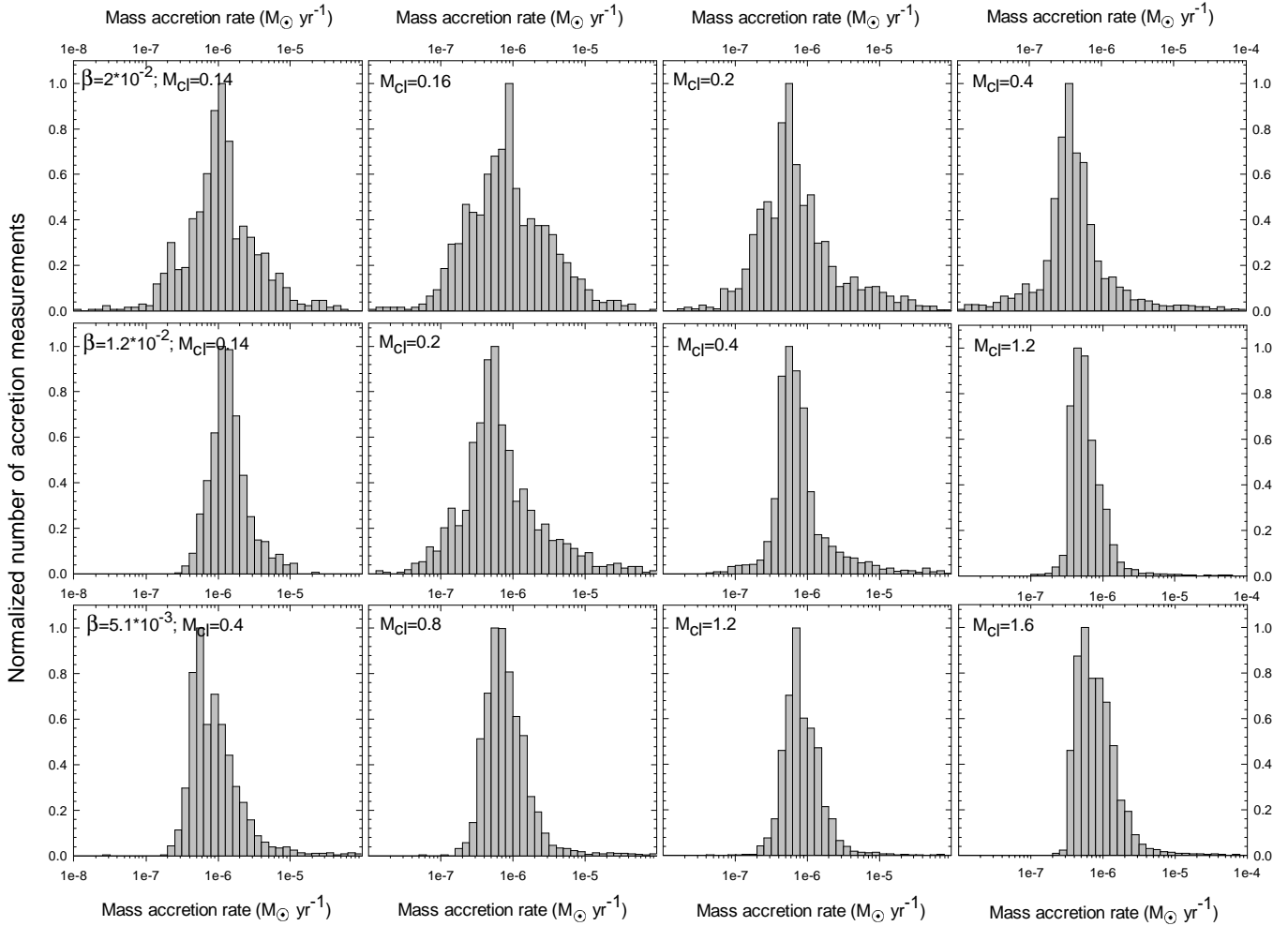}
      \caption{Distribution functions of accretion rates showing the normalized number of 
      accretion 
      measurements in the Class I stage for 12 models with the same parameters as in Figure~\ref{fig1}.}
         \label{fig3}
\end{figure*}

We start by constructing the distribution function (DF) of accretion rates,
which shows the normalized number of YSOs versus accretion rate using 
data obtained by 
\citet{Enoch09} in  Perseus, Serpens, and Ophiuchus star-forming regions.
We distribute the observed accretion rates in 20 logaritmically spaced bins 
between $10^{-8}~M_\odot$~yr$^{-1}$ and $10^{-4}~M_\odot$~yr$^{-1}$.
The normalization is done by first finding a bin that happens to have the maximum number 
of YSOs and then normalizing all bins to this maximum number. The resultant
normalized distribution is shown in Figure~\ref{fig2}. 
The dashed/solid lines show gaussian/log-normal fits to the observed data.
Neither gaussian nor log-normal functions yield good results, though the latter
function is more preferable than the former. There is a lack of well defined peak 
in the observed accretion rates,
which may to some extent be attributed to a small number of observed objects.
The log-normal fit yields the following relation between the normalized number of YSOs
$N_{\rm YSO}$ and observationally inferred accretion rates $\dot{M}$
\begin{equation}
N_{\rm YSO}=0.7 \exp \left( -  {1 \over 2} \left[ {\ln(\dot{M}/\dot{M}_0) \over 2.1} \right]^2 \right),
\end{equation}
where $\dot{M}_0=4\times 10^{-7}~M_\odot$~yr$^{-1}$ is the geometric mean accretion rate. 
According to \cite{Enoch09},
there are approximately 55\% of YSOs with sub-Shu accretion rates ($\dot{M}<10^{-6}$~yr$^{-1}$)
and about $5\%$ of YSOs with super-Shu rates ($\dot{M}>10^{-5}~M_\odot$~yr$^{-1}$).

One feature in Figure~\ref{fig2} is worth special attention---there is a sizeable portion 
of YSOs with accretion
rates smaller than $10^{-7}~M_\odot$~yr$^{-1}$. These objects are usually characterized by
bolometric luminosity $L_{\rm bol}<0.1~L_\odot$ and constitute a group of objects 
called VELLOs or very low luminosity objects \citep[e.g.][]{Young04,Kauffmann05,Bourke06, 
Stecklum07}. Out of 89 Class I YSOs in the compilation of \citet{Enoch09}, 19 objects are VELLOs,
which is roughly 21\%. 

To compare our models with observations, we calculate $\dot{M}$ every 20 yr in the 
Class I stage of stellar evolution.
In the following text, we refer to these calculations as ''accretion measurements'' or simply ''measurements''
by analogy to observations. We note that due to the use of the 
sink cell in our numerical simulations, the Class I stage in some models 
may start before the disk begins to form (see Fig.~\ref{fig1}). 
In such rare cases, accretion rates are calculated only starting from the disk formation epoch.
Next, we construct the distribution function of accretion rates showing 
the normalized number of accretion measurements versus the accretion rate 
in 50 logarithmically spaced bins between $10^{-8}~M_\odot$~yr$^{-1}$ and $10^{-4}~M_\odot$~yr$^{-1}$.
The normalization is done by first finding a bin that happens to have the maximum number 
of accretion measurements and then normalizing all bins to this maximum number.

Figure~\ref{fig3}
presents DFs in the Class I stage for the same set of models as in Figure~\ref{fig1}. Model 
accretion rates approximately follow a log-normal distribution, the geometric mean values are 
mostly located in the 0.5--1.0$\times 10^{-6}~M_\odot$~yr$^{-1}$ range. There is a 
mild tendency for models with larger values of $\beta$ to have a wider range of accretion rates.
It is also seen that the magnitude of intrinsic variability in most models is insufficient to account
for a wide spread in observed mass accretion rates (Fig.~\ref{fig2}), implying that
some object-to-object variations (in the initial cloud core masses and/or rotation rates) 
are needed to reproduce the observed spread. 

Another interesting feature in Figure~\ref{fig3} is a clear trend that, for the same $\beta$, 
the peak $\dot{M}$ decreases with increasing cloud core mass. This 
tendency is
especially apparent for cloud cores with sufficiently high rotation rates, 
$\beta\ga 1.2\times10^{-2}$ and can be explained by a growing strength of the burst mode of accretion
along the sequence of increasing cloud core masses. Indeed, Figure~\ref{fig1} demonstrates that, for
the same $\beta$, the
number of accretion bursts increases with increasing cloud core mass, which is explained by 
growing 
disk masses and increasing propensity of disks to fragmentation. At the
same time, periods of quiescent accretion between the bursts become also more 
frequent and are characterized by lower rates. This acts to decrease the peak $\dot{M}$ along 
the sequence of increasing
cloud core masses. By the same reason, the peak $\dot{M}$ has a weak dependence on $\beta$ for cloud
cores of similar mass.  

The fact that $\dot{M}$ in our numerical simulations are log-normally distributed is interesting. 
It suggests that the accretion process is not statistically independent but it has
a memory effect---short periods of elevated accretion are, as a rule, followed by
prolonged periods of quiescent accretion and vice versa. This pattern of time behaviour is not chaotic
but is rather controlled by disk physics, implying a causal link between 
accretion events that are separated not too far in time.

As the next step, we want to construct the integrated DF of accretion rates,
which would include data from all models in Table~\ref{table1}. 
Every individual model
is characterized by a distinct duration of the Class I stage, which increases
along the sequence of increasing stellar masses. This means that models with longer 
lifetimes of the Class I stage are more significant statistically and have a larger number of 
accretion measurements than models with shorter Class I lifetimes.  
At the same time, objects at the lower mass end
(with shorter Class I lifetimes) are expected to be more abundant. The letter
tendency, however, may be counterbalanced by accretion 
measurements biased toward more luminous (and massive) objects. Therefore, we first 
sum up non-normalized DFs for every models and then normalize the integrated DF
in the same manner as described above, thus accentuating models with a larger number  
of accretion measurements.

\begin{figure*}
 \centering
  \includegraphics[width=15cm]{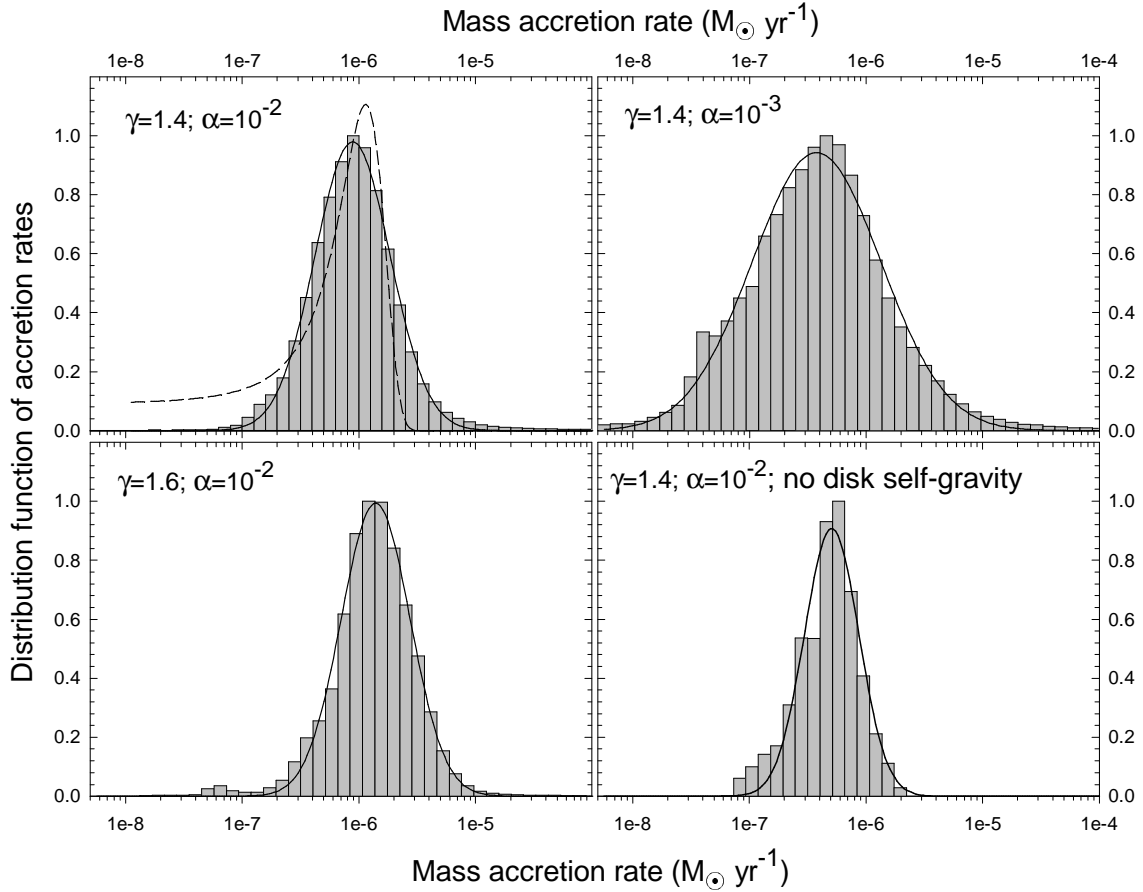}
      \caption{Integrated distribution function of accretion rates in the Class I stage showing
      the normalized number of accretion measurements $N_{\rm model}$ 
      versus mass accretion rate $\dot{M}$ for all models of Table~\ref{table1}. 
      Every panel is characterized by distinct values of the $\alpha$-parameter
      and the ratio of specific heats $\gamma$ as indicated. In addition,
      the lower-right panel has disk self-gravity artificially set to zero.
      The solid and dashed
      lines are the log-normal and gaussian fits to the model data, respectively.}
         \label{fig4}
\end{figure*}

The resultant integrated DF is presented in the upper-left 
panel of Figure~\ref{fig4}.
The solid line shows a log-normal fit to the model data, while the dashed line is a gaussian fit. 
It is evident that the log-normal function yields a much better fit to the model data 
than the gaussian one. In particular, we find the following relation between the normalized 
number of accretion 
measurements $N_{\rm model}$ and the mass accretion rates $\dot{M}$
\begin{equation}
N_{\rm model}=0.97 \exp \left( -  {1 \over 2} \left[ {\ln(\dot{M}/\dot{M}_0) \over 0.75} \right]^2 \right),
\end{equation}
where $\dot{M}_0=8.8\times 10^{-7}~M_\odot$~yr$^{-1}$ is the geometric mean value.
Our modeling predicts that about $45\pm 5\%$ of YSOs are expected to have ``sub-Shu'' 
accretion rates  with $\dot{M}\le 10^{-6}~M_\odot$~yr$^{-1}$. Approximately the same number 
of YSOs are expected to have accretion rates that are roughly consistent with the standard model 
$10^{-6}~M_\odot~\mathrm{yr}^{-1} \le \dot{M}\le 10^{-5}~M_\odot$~yr$^{-1}$, and only 2\% of YSOs 
have ``super-Shu'' accretion rates with $\dot{M}>10^{-5}~M_\odot$~yr$^{-1}$.
These numbers  are in fare agreement with what obtained by \citet{Enoch09}.
However, our numerical modeling predicts a much smaller
fraction of VELLOs with $\dot{M}<10^{-7}~M_\odot$~yr$^{-1}$, $\sim 1.0\%$, in contrast to
$\sim 21\%$ in \citet{Enoch09}. This is most likely related to the fact 
we have adopted too large a value for the $\alpha$-parameter, $\alpha=0.01$, thus overemphasizing
the effect of viscosity in the early disk evolution. It is known that
viscosity acts to suppress gravitational instability and associated large variations 
in the accretion rates \citep{VB09},
reducing the number and frequency of FU Orionis or VELLO events.
The effect of a smaller $\alpha$-parameter on our model mass accretion rates 
is discussed below.

%\begin{figure}
%  \resizebox{\hsize}{!}{\includegraphics{figure4.eps}}
%      \caption{Integrated distribution function of accretion rates in the Class I stage showing
%      the normalized number of accretion measurements $N_{\rm model}$ 
%      versus mass accretion rate $\dot{M}$ for all models of Table~\ref{table1}. 
%      The solid and dashed
%      lines are the log-normal and normal fit to the model data, respectively.}
%         \label{fig4}
%\end{figure}

\subsection{The effect of lower disk viscosity}
\label{lowervisc}
We have considered so far model disks with $\alpha=10^{-2}$. 
As discussed in Section~\ref{nummodel}, this value of the spatially and temporally averaged 
$\alpha$-parameter is  a realistic upper limit, though some short-living
fluctuations towards larger values are possible. In this section, we consider disks with an order of
magnitude lower $\alpha$-parameter. The upper-right panel in Figure~\ref{fig4} presents the integrated
distribution function of mass accretion rates for all models from Table~\ref{table1} with $\alpha$ set
to $10^{-3}$. The solid line is the log-normal fit to the model data described as
\begin{equation}
N_{\rm model}=0.94 \exp \left( -  {1 \over 2} \left[ {\ln(\dot{M}/\dot{M}_0) \over 1.25} \right]^2 \right),
\end{equation}
where $\dot{M}_0=3.8\times 10^{-7}~M_\odot$~yr$^{-1}$.

Numerical simulations with $\alpha=10^{-3}$
reveal a substantial number ($\sim 15\%$) of VELLOs 
with $\dot{M}\la 10^{-7}~M_\odot$~yr$^{-1}$, in fare agreement with
Figure~\ref{fig2} and data of \citet{Enoch09}. On the other hand, models with $\alpha=10^{-3}$
predict that about $74\pm 4\%$ of YSOs have sub-Shu accretion rates with 
$\dot{M}\le 10^{-6}~M_\odot$~yr$^{-1}$ and only $24\pm4\%$ of 
them have accretion rates that are consistent with the standard model 
$10^{-6}~M_\odot~\mathrm{yr}^{-1} \le \dot{M}\le 10^{-5}~M_\odot$~yr$^{-1}$,
in contrast to 55\% and 40\%, respectively, found by \citet{Enoch09}.
It is evident that the $\alpha=10^{-3}$ models somewhat overestimate the
number of sub-Shu objects as inferred from observations.  
This suggests that a preferable value for the temporally 
and spatially averaged $\alpha$-parameter may lie somewhere between 
$10^{-3}$ and $10^{-2}$.
In both limiting cases, about 2\% of YSOs are expected to have 
super-Shu accretion rates with $\dot{M}>10^{-5}~M_\odot$~yr$^{-1}$.

%\begin{figure}
%  \resizebox{\hsize}{!}{\includegraphics{figure3a.eps}}
%      \caption{The same as Figure~\ref{fig3} but only for $\alpha=10^{-3}$}
%         \label{fig3a}
%\end{figure}
%
%$\alpha\sim0.01$  As \citet{VB09} have recently demonstrated,

\subsection{The effect of higher disk temperature}
\label{hotdisk}
In this section, we briefly consider the effect that a greater ratio of specific heats $\gamma=1.6$
may have on the mass accretion rates. An increase in $\gamma$ in barotropic models leads to
an increase in disk temperature. For instance, in the $\gamma=1.4$ case 
a mean disk temperature at r=10~AU (in the Class I stage)
for nine models of Figure~\ref{fig1} is 29--32~K, while in the $\gamma=1.6$
case it equals to 45--50~K.
The corresponding integrated distribution function of accretion rates is shown in the lower-left panel
of Figure~\ref{fig4}. The $\alpha$-parameter is set to $10^{-2}$.
As usual, the solid line is the log-normal fit to the model data described by the following relation
\begin{equation}
N_{\rm model}=1.0 \exp \left( -  {1 \over 2} \left[ {\ln(\dot{M}/\dot{M}_0) \over 0.68} \right]^2 \right),
\end{equation}
where $\dot{M}_0=1.4\times 10^{-6}~M_\odot$~yr$^{-1}$.
 A visual comparison of the upper-left and lower-left panels 
demonstrates that the spread in $\dot{M}$ in the $\gamma=1.6$ models 
is narrower than in the $\gamma=1.4$ case, which is a direct consequence of
weaker gravitational instability in warmer disks.
%feature a smaller number of objects with the sub-Shu accretion rates ($\dot{M}\la
%10^{-6}~M_\odot$~yr$^{-1}$). 
More specifically, there are $37\%\pm5\%$ sub-Shu objects and $62\% \pm 5\%$ ``standard'' objects 
with $10^{-6}~M_\odot~\mathrm{yr}^{-1} \la \dot{M} \la 10^{-5}~M_\odot$~yr$^{-1}$, whereas
observational data indicate a prevalence of sub-Shu accretors ($\sim 55\%$).
The shortage of sub-Shu accretors in the $\gamma=1.6$ case as compared to \citet{Enoch09}
suggests that the observed circumstellar disks may be relatively cold. 
Alternatively, a combination of warmer disk temperature and lower turbulent viscosity 
(with $\alpha\sim 10^{-3}$) may also provide a better fit to \citet{Enoch09}.

\subsection{Purely viscous disks}
As the last numerical experiment, we consider circumstellar disks in which mass and 
angular momentum transport is governed exclusively by turbulent viscosity. This is done with 
the purpose to investigate the
role of disk self-gravity in the early embedded phase of stellar evolution. We artificially set 
disk self-gravity to zero immediately after disk formation, but the gravity of the central star
is kept unchanged. The resultant integrated DF of accretion rates for all models of Table~\ref{table1}
is shown in the lower-right panel of Figure~\ref{fig4}.
Purely viscous models greatly underestimate the number of objects with ``standard'' accretion 
rates $10^{-6}~M_\odot~\mathrm{yr}^{-1} \la \dot{M}\la 10^{-5}~M_\odot$~yr$^{-1}$.
In particular, modeling predicts $7\%$ of such objects, in contrast to $\sim 45\%$ in the sample of
\citet{Enoch09}. There are no objects with super-Shu accretion rates $\dot{M}\ga 10^{-5}~M_\odot$~yr$^{-1}$.
This example serves to demonstrate the importance of disk self-gravity in the early disk evolution.

\section{Accretion rates versus stellar masses}
It has recently been noticed that the mass accretion rates of TTSs and BDs of 0.5--3.0~Myr age
have a strong dependence on the central object mass $M_\ast$ of the form 
$\dot{M}\propto M_\ast^n$ with $n\approx2.0$
\citep[see e.g.][]{Muzerolle03,Natta04}. 
%More specifically, the mass accretion rate
%seems to scale with the object mass as $\dot{M}\propto M_\ast^n$, where $n\approx2.0$. 
A numerical model that is used in the present work can in principal reproduce this relation \citep{VB10},
predicting a somewhat steeper dependence for BDs and low-mass TTSs ($n\approx 2.9\pm 0.5$), as
observed, and
a shallower dependence for intermediate- and upper-mass TTSs ($n\approx 1.5\pm 0.1$), as also observed.
It is therefore very interesting to see if a similar relation is expected in the early embedded phase
of stellar evolution.

\begin{figure}
  \resizebox{\hsize}{!}{\includegraphics{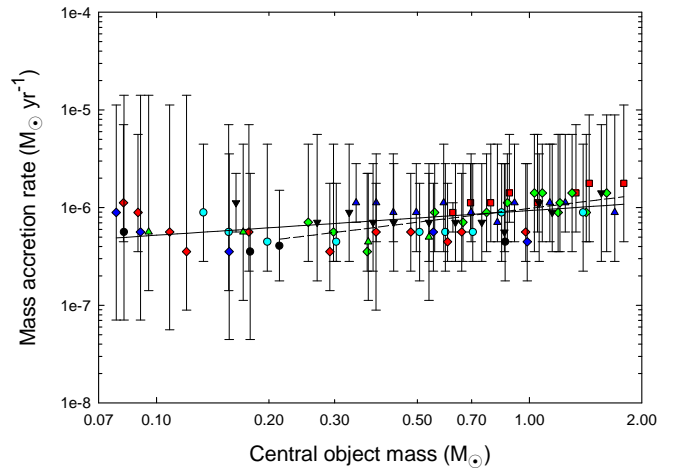}}
      \caption{Geometric mean mass accretion rates versus time-averaged stellar masses 
      in the Class I stage of stellar evolution. Symbols of specific shape and color represent
      different sets of models as described in the text (see online version for the color figure). 
      Vertical bars delineate typical variations
      in the mass accretion rate in every model. The lines are the least-squares 
      best fits to all data (solid) and intermediate- and upper-mass objects (dashed). }
         \label{fig5}
\end{figure}

For every model from Table~\ref{table1}, we calculate the geometric mean mass accretion rate
$\dot{M}_{\rm g.m.}$ as a value at which the corresponding DF of accretion rates
reaches a maximum. All models in this section have $\alpha$ and $\gamma$ set to $10^{-2}$ and 1.4, respectively.
Color symbols in Figure~\ref{fig5} present $\dot{M}_{\rm g.m.}$ (in $M_\odot$~yr$^{-1}$) versus 
time-averaged stellar masses $\langle M_{\ast} \rangle$ 
(in $M_\odot$) in the Class I stage of stellar evolution. 
The time-averaging is done over the duration of the Class I stage, which is distinct in each 
model. 
%$and is calculated separately for each model using the envelope mass as an indicator 
(%see Section~\ref{timeevol}).
In particular, model set 1 is plotted by red squares, model set 2---by blue triangles-up,
model set 3---by green diamonds, model set 4---by black triangles-down, 
model set 5---by cyan circles, 
model set 6---by red diamonds, model set 7---by green triangles up, model set 8---by blue diamonds,
and model set 9---by black circles. Each filled
symbol (of same color and shape) within a given set of models represents an individual 
object, which has formed from a cloud core of distinct mass, 
rotation rate, and size.

In addition, each filled symbol is assigned vertical 
bars describing (for a given model) a typical range of mass accretion 
rates in the Class I stage. Defining these typical values turned out not trivial.
This is because some of the models may show transient drops or rises in $\dot{M}$,
which are very short-lived and thus are not statistically significant.
Therefore, the vertical bars comprise only those accretion rates the DF for which is greater 
than 0.05.
%meaning that only those $\dot{M}$ 
%that have a detection probability greater than one part in twenty are shown in the figure.
This means that we have excluded objects with extreme accretion rates, which have probability
to be detected less than one part in twenty. Considering the number of detected objects 
along the line of constant central object mass ($\la 20$) in \citet{Enoch09}, 
we believe that our estimate represents conservative values. 
%To sum up,
%the error bars in Figure~\ref{fig2} should be regarded as a typical 
%range of accretion rates in the Class I phase

Figure~\ref{fig5} clearly demonstrates that the mass accretion rates in the embedded phase
do {\it not} show any strong dependence on the stellar mass. The least-squares best fit
to all data (solid line) yields the following relation
\begin{equation}
\dot{M}_{\rm g.m.}= 10^{-6} \langle M_\ast \rangle^{0.25\pm0.05},
\label{relation1}
\end{equation}
which is considerably shallower than that for the TTSs and BDs. 
Accretion rates of the intermediate- and upper-mass objects ($M_\ast\ga 0.2~M_\odot$) 
in Figure~\ref{fig5} seem to have a somewhat steeper dependence on $\langle M_\ast \rangle$,
as indicated by the dashed line, yet the corresponding exponent $0.5\pm0.1$ is considerably smaller
than that of the evolved TTSs and BDs of 0.5-3.0~Myr age. 
We again want to emphasize that our model does reproduce a steep dependence
seen in TTSs and BDs for both purely self-gravitating disks \citep{VB08} and self-gravitating 
disks with turbulent viscosity \citep{VB10}.

A hint as to why this difference takes place can be seen in Figure~\ref{fig1}.
The embedded phase is characterized by mass infall rates onto the disk $\dot{M}_{\rm infall}$ 
(dashed lines) that are often greater than the mass accretion rate onto the star (solid lines).
This implies that accretion in the embedded phase may in part be controlled
by the rate of gas deposition onto the disk, which in turn is expected to be only 
weakly dependent on the stellar mass. 

\begin{figure}
  \resizebox{\hsize}{!}{\includegraphics{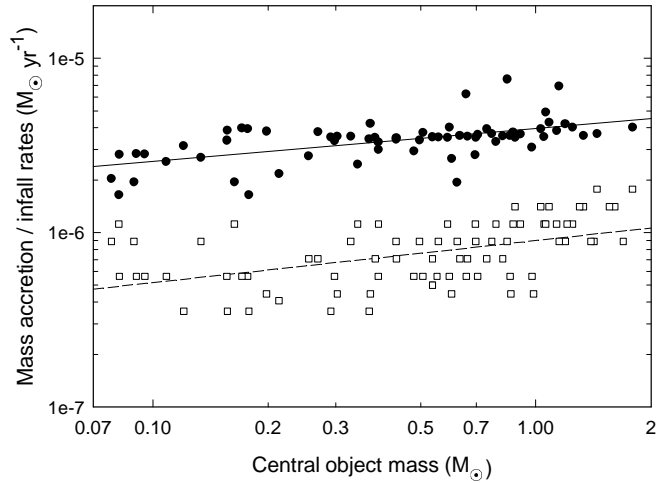}}
      \caption{Geometric mean mass accretion rates onto the star $\dot{M}_{\rm g.m.}$ 
      (open squares) 
      and time-averaged mass infall rates onto the disk $\langle \dot{M}_{\rm infall} \rangle$ 
      (filled circles) versus time-averaged stellar masses in the 
      Class I stage of stellar evolution. Most models from Table~1 are shown in the figure.
      The dashed and solid lines are the least squares best fits to $\dot{M}_{\rm g.m.}$ 
      and $\langle \dot{M}_{\rm infall} \rangle$, respectively. }
         \label{fig6}
\end{figure}

We demonstrate this in Figure~\ref{fig6}, which
shows the time-averaged infall rates $\langle \dot{M}_{\rm infall}\rangle$ (filled circles) 
versus time-averaged stellar masses $\langle M_\ast \rangle$ for most models in Table~\ref{table1}.
We have eliminated some of the models, which happen to
have disk radii that exceed 600 AU---the radius at which we calculate the mass infall
rate. The time-averaging is done over the duration of the Class I stage. 
For comparison, we also plot the geometric mean mass accretion rates $\overline{\dot{M}}$ 
(open circles). Note that we use the same symbol type for all models in this figure.
It is evident that $\langle \dot{M}_{\rm infall} \rangle$ is {\it systematically} greater 
than $\dot{M}_{\rm g.m.}$. The least squares fit 
yields the following relation between the time-averaged infall rates and stellar masses
\begin{equation}
\langle \dot{M}_{\rm infall} \rangle= 4.0\times 10^{-6} \langle M_\ast \rangle^{0.20\pm0.03},
\end{equation}
the exponent of which is quite similar to that of the $\dot{M}_{\rm g.m.}$ vs. $\langle M_\ast \rangle$
relation (dashed line, see also equation~[\ref{relation1}]). 
A weak dependence of $\langle \dot{M}_{\rm infall} \rangle$ on the stellar mass is 
explained by a gradual increase in the envelope temperature
along the sequence of increasing stellar masses.
We also emphasize that our model infall rates lie within the limits predicted by the standard model
of \citet{Shu77} for star forming regions with gas temperature $T_{\rm g}=10-25$~K.

%The dashed line shows the 

\section{Conclusions}
In this paper, we have studied mass accretion rates on to the star $\dot{M}$ 
and mass infall rates onto the disk $\dot{M}_{\rm infall}$ in the early embedded phase of
stellar evolution, focusing on objects with stellar masses lying in the 0.08--2.0~$M_\odot$ range.
We have employed numerical hydrodynamic simulations of the gravitational collapse  
of a large number of model cloud cores, which cover a wide range of masses, initial rotation rates,
and sizes. The numerical integration is started in the {\it pre-stellar} phase and 
followed into the late accretion phase when almost all of the
initial cloud core has been accreted onto the star/disk system. We compare our model
accretion rates with those recently inferred from observations of embedded YSOs in young
star forming regions. We find the following.

\begin{itemize}

\item Our numerical modeling yields a highly variable accretion history in the phase that 
immediately follows disk formation. Prolonged periods of low accretion 
$ \dot{M} \la 10^{-6}~M_\odot$~yr$^{-1}$,
when disks may be gravitationally unstable but are yet stable to fragmentation,
are interspersed with short-lived bursts of mass accretion 
with $\dot{M}\ga 10^{-5}~M_\odot$~yr$^{-1}$, when
disks forms fragments and drive them onto the central star via the exchange of angular
momentum with spiral arms. The latter phenomenon is called the burst mode of accretion
and is investigated in detail in \citet{VB06}.
%as implied by observations of \citet{Enoch09}. This variability is caused by ....

\item Model accretion rates $\dot{M}$ in the early embedded phase have a log-normal distribution, with
its shape depending on the magnitude of turbulent viscosity (parameterized
using the usual $\alpha$-model of \citet{SS}) and disk temperature.

\item The spread in $\dot{M}$ (or the width of the accretion rate distribution function) 
is greater in models with lower $\alpha$
and smaller in models with higher $\alpha$ and higher disk temperature $T_{\rm d}$. 
An increase in either $T_{\rm d}$ or $\alpha$ (or both) acts to stabilize 
disk against gravitational instability \citep{VB09}. This suggests that 
gravitational instability  may be a dominant cause of
the accretion diversity in circumstellar disks exposed to
a continuing gas deposition from infalling envelopes.

\item Our models yield a large population of objects (37\%--75\%) 
with mass accretion rates $\dot{M}\la 10^{-6}~M_\odot$~yr$^{-1}$, which are smaller than those 
predicted by the standard model of inside-out collapse,
$\mathrm{a~few}\times 10^{-6}~M_\odot$~yr$^{-1}$ \citep{Shu77}.
A similar large fraction of embedded YSOs 
with sub-Shu accretion rates ($\sim 55\%$) was recently reported by \citet{Enoch09}
for Perseus, Serpens, and Ophiuchus star forming regions. 

%\item Models with low turbulent viscosity $\alpha=10^{-3}$ tend to 
%overestimate the number of observed sub-Shu YSOs, while models with $\alpha=10^{-2}$
%underestimate this number. Models with increased 
%disk temperatures tend to yield a smaller number of sub-Shu objects,
%suggesting that gravitational instability 

\item Approximately 1\%--2\% of objects in our modeling have 
super-Shu accretion rates $\dot{M}>10^{-5}~M_\odot$~yr$^{-1}$, 
which is roughly a factor of 2--3 smaller than found by \citet{Enoch09}.
This suggest that other (than disk fragmentation) mechanisms may also 
contribute to bursts of mass accretion in the embedded phase stellar evolution.

\item Purely viscous models, with disk self-gravity artificially set to zero,
significantly underestimate the number of objects with ``standard'' mass accretion
rates $10^{-6}~M_\odot~\mathrm{yr}^{-1} \la \dot{M}\la 10^{-5}~M_\odot$~yr$^{-1}$, yielding essentially no objects with super-Shu rates. This demonstrates
the importance of disk self-gravity in the early disk evolution.  

\item Mean mass accretion rates  in the early embedded phase 
show only a weak dependence on the time-averaged stellar masses
of the form $\dot{M}_{\rm g.m.}=10^{-6}\langle M_\ast \rangle^{0.25\pm0.05}$,
in sharp contrast to the steep (with exponent $\sim 2.0$) empirical relation inferred 
for evolved TTSs and BDs of 0.5--3.0~Myr age 
\citep[see e.g.][]{Muzerolle03,Natta04}. This result is particularly interesting
since our numerical model can reproduce the steep relation in the late evolution phase
for both purely self-gravitating disks \citep{VB08} and self-gravitating disks with 
turbulent viscosity \citep{VB10}.

%yields a steep $\dot{M}_{\rm g.m}$--$\langle M_\ast\rangle$ 
%relation, as observed, for evolved TTSs and BDs \citep{VB08,VB10}.

\item The lack of strong dependence of $\dot{M}_{\rm g.m}$ on $\langle M_\ast \rangle$ 
in the early 
embedded phase can be attributed to a continuing deposition of matter from the 
infalling envelope. The time-averaged rate of mass infall onto the disk 
$\langle \dot{M}_{\rm infall} \rangle$ is systematically greater (roughly by a
factor of four) than the mean mass accretion rate onto 
the star $\dot{M}_{\rm g.m.}$, 
implying that $\dot{M}_{\rm g.m.}$ is at list partly determined by 
$\langle \dot{M}_{\rm infall} \rangle$.
At the same time, $\langle \dot{M}_{\rm infall} \rangle$ exhibits only a week growth 
along the sequence of increasing stellar masses, which is caused by a moderate 
increase in the envelope temperature from $\sim 10$~K to $\sim 25$~K.

\end{itemize}

\acknowledgments
   The author gratefully thanks the anonymous referee for a helpful report
   and Prof. Shantanu Basu for stimulating discussions. 
   The author gratefully acknowledges present support 
   from an ACEnet Fellowship. Numerical simulations were done 
   on the Atlantic Computational  Excellence Network (ACEnet).

\end{document}